\documentclass[11pt]{article}

\usepackage{amssymb,amsfonts,latexsym,amscd,amsthm}
\usepackage{color,epsfig,times}
\usepackage[colorlinks=true,plainpages=false,linktocpage,linkcolor=blue,citecolor=green,pagebackref]{hyperref}
\usepackage[letterpaper,left=1.5in,top=1in,bottom=1in,right=1.5in,includeheadfoot,headheight=16.6pt]{geometry}
\usepackage{graphicx}

\usepackage{amsmath}

\begin{document}

\title{Nash equilibrium quantum states and optimal quantum data classification }
\author{Faisal Shah Khan\footnote{Department of Applied Mathematics and Sciences, Khalifa University UAE, Email: faisal.khan@kustar.ac.ae}}
 %\hspace{2mm} Simon J.D.Phoenix\footnote{Department of Applied Mathematics and Sciences, Khalifa University, Email: simon.phoenix@kustar.ac.ae}, \hspace{2mm} Ahmed El Hady\footnote{Department of Non-linear Dynamics and Network Dynamics Group, Max Planck Institute for Dynamics and Self Organization, Email:ahmed@nld.ds.mpg.de}}
%\author{Faisal Shah Khan\footnote{Department of Applied Mathematics and Sciences, Khalifa University, Email: faisal.khan@kustar.ac.ae}, \hspace{2mm} Simon J.D.Phoenix\footnote{Department of Applied Mathematics and Sciences, Khalifa University, Email: simon.phoenix@kustar.ac.ae}, \hspace{2mm} Ahmed El Hady\footnote{Department of Non-linear Dynamics and Network Dynamics Group, Max Planck Institute for Dynamics and Self Organization, Email:ahmed@nld.ds.mpg.de}}
\maketitle

\begin{abstract}
This letter reports a novel application of game theory to quantum informational processes which can be used to optimally classify data generated by these processes. To this end, the notion of simultaneously distinguishing a pure quantum state, generated by a quantum informational process, from its constituent observable states optimally - given the constraint of these observables being orthogonal to each other, is first introduced. This problem is solved via a non-cooperative game model and the affiliated solution concept of Nash equilibrium. The notion of Nash equilibrium quantum states is introduced and used to classify quantum data optimally.
\end{abstract}


\section{Introduction}

In its manifestation as a higher order of information processing via quantum superposition followed by quantum measurements instead of probability distributions, quantum mechanics offers several advantages over classical information processing which can be harnessed for technological innovations, such as quantum computers that sometimes dramatically surpass the computational efficiency of classical computers. Other innovations based on this quantum information perspective include communication technologies such as quantum key distribution protocols for provably secure encryption of data, and imaging protocols which can circumvent the catch-22 of image processing where higher image resolution demands higher energy, destructive probing wavelengths. 

Underlying all these quantum technological innovations is the idea of quantum data and its control. Quantum data is simply digits of quantum information, or qudits, an idea further elaborated in the following section. Control of these qudits is performing quantum mechanical actions on them so as to get them in some desired configuration. The configuration of qudits of interest here is one where they can be optimally classified with respect to the elements of a given set of observables. In other words, is it possible, using quantum mechanical operations, to group qudits between a given observable state $b$ and another quantum state $q$ such that the distance between $b$ and $q$ is as large as possible, for all observables in the given set? An answer to this question is proposed here using a non-cooperative game model and the corresponding optimality concept of Nash equilibrium.

\section{The quantum system}

Recall that the mathematical model for quantum objects with finitely many observable states is a (projective) complex, finite $d$-dimensional Hilbert space ${\bf H}_d$. The elements of an orthogonal basis $B$ of the quantum system ${\bf H}_d$ represent the physically observable states of the quantum object. General linear combinations of the elements of $B$, known as quantum superpositions, while equally meaningful physically, cannot be observed directly; instead quantum superpositions can be observed to be in one of their constituent observable states with a certain probability via the physical process of (quantum) measurement. To be more precise, measurement is orthogonal projection of a quantum superposition $q$ onto its constituent observable states in $B$, with the length of each projection representing the probability with which $q$ is observed to be in the corresponding observable state. The following constrained optimization question arises naturally here: is there an optimal measurement of $q$? 

The constraints in this problem arise from the orthogonality of the observable states, that is, increasing or decreasing the probability of observing $q$ in any one observable state necessarily effects the probability of observing $q$ in at least another observable state. In particular, decreasing the probability of observing $q$ in a given observable state $b$ so as to distinguish $q$ from $b$ necessarily makes $q$ less distinguishable from some other observable states. Hence, the question of optimal measurement for $q$ can be cast as a problem of optimal distinguishably of $q$ from its constituents observable states $b_i \in B$, given the constraints of the orthogonality of the $b_i$. This is the problem of optimal simultaneous distinguishability of $q$ from the $b_i$. 

To elaborate further, note that the probability of observing $q$ in any one $b_i$ can be expressed as a function of the inner-product $\langle , \rangle$ of ${\bf H}_d$ and therefore as a function of the angle $\theta_{(q,b_i)}$ between $q$ and $b_i$. Further note that this functional relationship has a diametrically opposite nature so that whenever $\theta_{(q,b_i)}$ is small, the probability that $q$ will be observed to be in the state $b_i$, $|\langle q , b_i\rangle|$, is large, and vice versa. Now, the problem of optimal simultaneous distinguishability of $q$ from the $b_i$ is reduced to simultaneous minimization of the quantities $\theta_{(q,b_i)}$ for all $i$, given the orthogonality constraints $\langle b_i, b_j \rangle=0$ for all $i \neq j$. 

This problem begins to take on a game-theoretic flavor when one considers the basis elements $b_i$ to be the payoffs for some notion of ``players,'' and that these players have non-identical preferences over the various $b_i$. The constraints $\langle b_i, b_j \rangle=0$ for all $i \neq j$ now simply mean that if two players entertain non-identical preferences over $b_i$ and $b_j$, then these players will compete over distinguishing $q$ from these two basis states respectively. This notion of competition is one contribution of the game-theoretic flavoring just added, together with the exact mechanism for implementing this competition, namely, a non-cooperative game. Yet  another contribution that comes from game theory here is the notion of strategic choices made by the players in the game to realize an outcome that optimizes the distinguishability of $q$ from distinct $b_i$ and $b_j$. Summarizing, game theory offers the correct formalism for controlling the quantum physical problem at hand towards an optimal solution. Details appear in the following sections. 

%In the following section, the notion of a quantum game \cite{} is recalled to study the problem of optimal simultaneous distinguishability of a $q$ from all of the $b_i$, with the goal of obtaining Nash equilibrium solutions from the quantum game model. This quantum game will further be interpreted as a quantum data generator in section... and the corresponding Nash equilibrium solutions for the simultaneous distinguishability problem will be considered to be optimal quantum data classifiers. 

\section{Gaming the quantum system }\label{gaming the quantum}

An $n$ player, non-cooperative {\it quantum game} \cite{Khan} is a function $G$ with a finite-dimensional complex (projective) Hilbert space ${\bf  H}_d$ of quantum superpositions as its co-domain, combined with the additional feature of  ``players'' who entertain non-identical preferences over the elements of the co-domain. In symbols 
\begin{equation}\label{eq:equ1}
G : \Pi_i D_i \rightarrow {\bf H}_d, \quad i =1, 2, \dots, n.
\end{equation}
The factor $D_i$ in the domain of $G$ is the set of strategies of player $i$, and a play of the game $G$ is a tuple of strategies in $\Pi_i D_i$ producing a payoff to each player in the
form of an outcome, that is, an element of ${\bf H}_d$. 

A {\it Nash equilibrium} is a play of $G$ in which every player employs a strategy that is a best reply, with respects to his preferences over the outcomes, to the strategic choice of every other player. In other words, unilateral deviation from a Nash equilibrium by a player in the form of a different choice of strategy will produce an outcome which is less than or equal to in preference to that player than before. The quantum state that is the image of a Nash equilibrium play of $G$ is called a {\it Nash equilibrium quantum state}.  

The problem of optimal simultaneous distinguishability of $q$ from the $b_i$ now has a game-theoretic solution; that is, $q$ is optimally distinguishable from all the $b_i$ if $q$ is a Nash equilibrium quantum state. Note that gaming the quantum system ${\bf H}_d$ gives rise to the capability of controlling this problem towards an optimal solution in a very general sense. 
That is, the quantum game $G$ and the strategy sets $D_i$ can be very general mechanisms and therefore solutions to the problem can be constructed in several context of both mathematical and physical interests. 
%One context of quantum physical interest would be quantum computational. Optimal solutions for $q$ in this context have been provided in \cite{} where two qubit quantum computations have been modeled as a strictly competitive quantum game and as the non-strictly competitive game known as Prisoner's Dilemma, respectively. 
However, the exact nature of the control mechanism (quantum game) determines the exact nature of the optimal solution to the problem. This game model dependency of the solution is discussed in the following section with the aid of a particular game model. 

\subsection{Nash equilibrium quantum states}\label{NE qstates}

Let 
$$
B=\left\{ b_1, \dots, b_d \right\} 
$$
be an orthonormal basis (set of observable states) in ${\bf H}_d$ and let 
\begin{equation}\label{eqn:pref}
{\rm Pref}_k: b^{k}_{j_1} \succ b^{k}_{j_2} \succ \dots \succ b^{k}_{j_d}, \quad k=1,2, \dots, n
\end{equation}
be the preference profile of Player $k$ over the elements of $B$, with the symbol $\succ$ representing the notion of ``prefers over''. Hence, player $k$ prefers the element $b^{k}_{j_1}$ over the the element $b^{k}_{j_2}$ of $B$, and so forth in ascending order of the lowest subscript, until all the elements of $B$ are exhausted. The quantum fidelity \cite{Uhlmann} equation
\begin{equation}\label{eq:angle}
\cos \theta_{(p,q)}=|\left \langle p, q \right \rangle|
\end{equation}
for arbitrary elements $p$ and $q$ of ${\bf H}_d$ allows defining the players' preferences over arbitrary elements of ${\bf H}_d$ as follows. Player $k$ will prefer an arbitrary element $p$ of ${\bf H}_d$, that is, a quantum superposition of the elements of $B$, over another $q$, whenever $p$ is ``closer'' to {\it all} the elements of $B$ {\it with respect to} ${\rm Pref}_k$ than $q$ is. The notion of closeness utilized in the preceding sentence is the one described by the quantity $||p- q ||$ and which is an increasing function of $\theta_{(p,q)}$. The notion of players' preference over arbitrary quantum states in ${\bf H}_d$ is expressed symbolically as
\begin{equation}\label{eqn:arbpref}
p \succ q \quad {\rm whenever} \quad \theta_{(p,b^{k}_{j_i})} < \theta_{(q,b^{k}_{j_i})}, \forall j_i, \forall k
\end{equation}
and can be used to characterize the notion of Nash equilibrium in the non-cooperative quantum game $G$.

Suppose $E^*=(e_1^*, e_2^*, \dots, e_k^*, \dots, e_n^*)$ is a Nash equilibrium play in the non-cooperative quantum game $G$, producing the Nash equilibrium quantum state $G(E^*) \in {\bf H}_d$. Then a unilateral deviation from $E^*$ by Player $k$ to some other play $E=(e_1^*,e_2^*, \dots, e_k, \dots e_n^*)$ will produce a quantum state $G(E)$ that will be less than or equal to in preference to Player $k$, with respect to ${\rm Pref}_k$, than before. Stated explicitly in terms of the quantity $\theta_{(,)}$, a Nash equilibrium play $E^*$ of the quantum game $G$ will satisfy the following inequalities:
\begin{equation}\label{NE}
\theta_{(G(E),b^k_{j_i})} \geq \theta_{(G(E^*),b^k_{j_i})}, \forall j_i, \forall k.
\end{equation}

The existence question of $E^*$ in $G$ can addressed by referring to the theory of Hilbert space \cite{Roman} if $G$ is a unitary map into ${\bf H}_{d}$, for in this case its image, ${\rm Im}(G)$, is a sub-Hilbert space of ${\bf H}_{d}$. Therefore, for every element $h \in {\bf H}_{d}$, there exist an element $k \in {\rm Im}(G)$ such that for all other $k{^\prime} \in {\rm Im}(G)$
\begin{equation}\label{Hilbspace}
\theta_{(k^{\prime},h)} \geq \theta_{(k,h)}.
\end{equation}
Setting 
$$
k^{\prime}=G(E); \quad k=G(E^*); \quad h=b_{j_i}^k
$$
in (\ref{Hilbspace}) and adding the universal quantifiers $\forall j_i$ and $\forall k$ recovers the Nash equilibrium condition in (\ref{NE}), thus establishing the necessary and sufficient conditions for the existence of a Nash equilibrium in a {\it unitary} non-cooperative quantum game $G$. In the more general setting where $G$ is not a unitary operation, the theory of Hilbert space requires that ${\rm Im}(G)$ at least be a complete convex subset of ${\bf H}_d$ in order to meet the existence conditions for Nash equilibrium quantum states characterized above. 

Further generalization can see the finite-dimensional Hilbert space ${\bf H}_d$ replaced with a Hilbert space that entertains a more general notion of observables, and appropriate mathematical conditions imposed to define the notion of Nash equilibrium in quantum states and explore their existence. 

\section{Optimal quantum data classification}

Note that a Nash equilibrium quantum state groups qudits in ${\bf H}_d$ according to their distance from the observable states $b_i$ in $B$, with respect to the preferences of {\it all} the players' preferences over the various $b_i$. The latter property makes these groupings optimal in the sense that they are as large as possible (or as small as possible, depending on the player's point of view).  A qudit $q$ can now be said to be optimally grouped between any specified $b_i$ and a Nash equilibrium state $G(E^*)$ and therefore can be classified as being of ``type'' $b_i$. Quantum states that are in the image of the quantum game $G$ or in other words, are generated by the quantum information process $G$, can now be classified optimally by type, as above, via Nash equilibrium states. 

Note that this classification of $q$ as being of the observable $b_i$ type does not guarantee that it will be measured as $b_i$ with a high probability, for the angle $\theta_{(q,b_i)}$ (or $\theta_{(G(E^*),b_i)}$ for that matter) might be such that the probability of $q$ measuring as $b_i$ is quite small due to the constraints of both the orthogonality of the $b_i$ and the non-identical nature of the players' preferences. Nonetheless, this classification is the best possible in terms of what can be considered as an optimal measurement of $q$ given the constraints here from quantum mechanics and game theory. And this may be enough for the purpose of the quantum information processes in many cases. For example, when it is the case that $\theta_{(G(E^*),b_i)}$ is such that the probability of measuring $q$ as $b_i$ is high (above 50$\%$), then this optimal classification scheme gives useful insights in the quantum informational process being modeled as a quantum game. For instance, if the quantum informational process $G$ is a quantum computation, and the desired outcome of the computation upon measurement is $b_i$, making $b_i$ the most preferred observable, then it is useful to know whether the quantum computation in fact entertains a Nash equilibrium quantum state, and that it has inputs for which the corresponding outputs $q$ are optimally classified with respect to $G(E^*)$ and $b_i$ and for which $\theta_{(G(E^*),b_i)}$ is as small as possible so that $q$ measures as $b_i$ with as high a probability as possible. Such insights make the implementation of a quantum computation worthwhile, given the costly endeavor such implementations are at the current stage of quantum technologies. 

It is important to keep in mind that the preferences of the players defined in (\ref{eqn:pref}) are just one way to define preferences; many others are possible, and therefore many other optimal classification schemes for qudits are possible. 
As an example, consider the optimal classification of qubits via Nash equilibrium quantum states using a strictly competitive quantum game with ${\bf H}_4$ where
$$
B=\left\{ b_1, b_2, b_3 b_4 \right\} 
$$
as a set of observable states entertaining the following preferences of the players: 
\begin{equation}\label{eqn:preftwo}
{\rm Pref}_1: b_1 \succ b_{2} \equiv_{3} \equiv b_4
\end{equation}
and
\begin{equation}\label{eqn:preftwo2}
{\rm Pref}_2: b_2 \succ b_{1} \equiv b_2 \equiv b_1.
\end{equation}
Here, the symbol $\equiv$ means ``indifferent between''. A Nash equilibrium quantum state $G(E^*)$ in this example will satisfy
$$
\theta_{(G(E),b_i)} \geq \theta_{(G(E^*),b_i)}
$$
with respect to both ${\rm Pref}_1$ and ${\rm Pref}_2$ and for any other quantum state $G(E)$. Because of the nature of the preferences of the players, the Nash equilibrium state classifies qubits in ${\bf H}_4$ into two groups; those that are of the type $b_1$ and those of the type $b_2$ since the players are indifferent between all other $b_i$. This game model was studied in \cite{Khan1} to gain insights of the kind mentioned in the preceding paragraph into two qubit quantum computations. 

\section{Discussion}

The main contribution of this paper is to show the usefulness of game theoretic reasoning in quantum information processing, particularly in light of the technological realization of such processes in the form of quantum communication protocols and quantum computers. The optimal classification of quantum data that comes about from game theoretic reasoning gives the capability to make informed decisions about the cost effectiveness of practically implementing expensive quantum information processes. Connections to other fields such as quantum state distinguishability or quantum state estimation are certainly conceivable, but are not considered to be of relevance to the main point of this letter. 

\section{Acknowledgment}
I gratefully acknowledge useful conversation with Ahmed El Hady. Thank you to the Aspen Center for Physics and the NSF Grant number $1066293$ for hospitality during the writing of this paper.

\end{document}